\documentclass[a4paper,11pt]{article}
\usepackage{jcappub}
\usepackage{multirow}
\usepackage{ulem}
\usepackage{graphicx} 

\title{A natural explanation of the Galactic Magnetic Fields from multistate Scalar Field Dark Matter}

\author[a,1]{Maribel Hern\'andez-M\'arquez,\note{Corresponding author.}}
\affiliation[a]{Instituto de Ciencias Nucleares, Universidad Nacional Aut\'onoma de M\'exico, Circuito Exterior C.U., A.P. 70-543, Ciudad de M\'exico 04510, M\'exico}

\author[b]{Bryan Mendoza-Meza,}
\affiliation[b]{Departamento de F\'{\i}sica, Centro de Investigaci\'on y de Estudios Avanzados del IPN, AP 14-740, Ciudad de M\'exico 07000, M\'exico}

\author[b,2]{Tonatiuh Matos,\note{Part of the Instituto
       Avanzado de Cosmolog\'{\i}a (IAC) Collaboration.}}

\author[c]{Tula Bernal,}
\affiliation[c]{\'Area de F\'isica, Depto.~de Preparatoria Agr\'icola, Universidad Aut\'onoma Chapingo, km~38.5 Carretera M\'exico-Texcoco, Texcoco 56230, Edo. M\'ex., M\'exico} 

\author[a]{and Miguel Alcubierre}

\emailAdd{maribel.hernandez@nucleares.unam.mx, luis.mendoza@cinvestav.mx, tonatiuh.matos@cinvestav.mx, tbernalm@chapingo.mx, malcubi@nucleares.unam.mx}

\abstract
{
In this article, we investigate the possibility that the large-scale magnetic fields observed in galaxies, of the order of microgauss, arise naturally from a complex Scalar Field Dark Matter (SFDM) halo charged under a local $U(1)$ symmetry. Extending our previous work, where multistate SFDM solutions were shown to form ``gravitational atoms'' capable of explaining the anisotropic distribution of satellite galaxies (VPOS), we analyze here the coupled dynamics of the scalar and a gauge field at the perturbative level. By solving the perturbed Klein-Gordon and gauge-field equations, we find the temporal evolution and show that the spatial structure of the induced electromagnetic fields is governed by the same spherical Bessel functions and spherical harmonics that characterize the ground and excited states of the multistate SFDM halo. Remarkably, the presence of the gauge field does not modify the dark-matter density distribution, which preserves the multistate configuration previously obtained. Our results demonstrate that a charged multistate SFDM halo can generate coherent, large-scale magnetic fields whose morphology is determined by the excited modes of the scalar field, providing a unified framework in which both galactic magnetic fields and VPOS-like structures originate from the underlying quantum nature of dark matter.
}

\date{\today}

\begin{document}

\maketitle

\flushbottom

\section{Introduction}
Dark matter is one of the most important problems currently existing in cosmology, and has been attempted to be solved in various ways since its discovery by Fritz Zwicky in 1933. He discovered that the observed matter cannot explain the rotation curves of galaxies, then there should be an invisible matter that could generate the necessary gravitational force to keep the stars in galaxies together.
The Cold Dark Matter (CDM) model was one of the first successful frameworks developed to address this issue; however, its predictions fail to match certain galactic-scale observations, such as the well-known core–cusp discrepancy and the missing satellites problem \cite{DelPopolo:2016emo}.
As an alternative to CDM, and as a possible solution to these issues, Scalar Field Dark Matter (SFDM) was proposed in 1998 \cite{guzman2000scalar} and has been referred in the literature with various names as Fuzzy \cite{hu2000fuzzy}, Bose-Einstein \cite{boehmer2007can} or Wave DM \cite{schive2014cosmic}. This hypothesis assumes that dark matter is a real scalar field with an ultralight mass, $m_{\Phi}\sim 10 ^{-22} eV$. This model has been studied for several years and has been able to explain observations such as the number of satellite galaxies around large galaxies \citep{Matos:2023usa}.

On the other hand, magnetic fields are present in all the Universe at different scales, from planets, to stars, to galaxies, to filaments and voids \cite{beck2013magnetic}. In the last years, coherent magnetic fields on scales of the order of $kpc$ have been observed in galaxies. These magnetic fields follow the spiral arms of galaxies, and their magnitude strength is of the order of microgauss \cite[see e.g.][]{beck2016magnetic}. The origin of these magnetic fields is a mystery, and there are different hypotheses that try to explain it. There are two broadly accepted theories on their origin: the first one is that the seed fields of magnetic fields were created in the early universe, these are known as primordial magnetic fields; and the other possibility is that the process of generation of the seed fields accompanies the gravitational collapse leading to structure formation \cite{durrer2013cosmological}. In both cases, the seed magnetic fields must be amplified through some dynamo mechanism in order to reach the present-day strengths observed in galaxies.

In order to explain these large-scale magnetic fields, in the present work we follow an alternative idea first proposed in \cite{hernandez2019could}. In that work, we proposed that dark matter is a complex scalar field charged under a local $U(1)$ symmetry, gauge invariant, requiring the introduction of a gauge boson $B_{\mu}$. This gauge boson could be a dark photon that only interacts with the photon of the standard model through a mixing term $\delta$ in the Lagrangian, or could be the photon of the standard model for an ultralight charge $q$, to be compatible with observations. In \cite{hernandez2019could}, the presence of a magnetic field induced by the gauge field of the scalar field was studied. It was shown that the presence of this gauge field could explain the large-scale magnetic fields observed in galaxies, with no modification of the rotation curves by the presence of such fields. In such work, an ansatz for the gauge field of the scalar field was used.

More recently, it has been shown that SFDM at finite temperature can form multistate configurations in which the halo occupies a superposition of ground and excited states. In particular, in \cite{bernal2025natural,matos2022quantum} it was demonstrated that these multistate halos behave as ``gravitational atoms'', with density profiles described by spherical Bessel functions and spherical harmonics. The first excited state exhibits a two-lobe structure aligned along the polar direction, which provides a natural explanation for the anisotropic distribution of satellite galaxies observed around the Milky Way (the VPOS phenomenon), Andromeda, Centaurus A, and other nearby galaxies \citep[see e.g.][and references therein]{bernal2025natural}. This result highlights the intrinsically quantum, anisotropic structure that SFDM halos can acquire.

Furthermore, in the Milky Way, measurements based on Faraday rotation of pulsars and extragalactic radio sources, together with synchrotron emission, indicate a total magnetic-field strength of approximately $5-10 \ \mu$G in the solar neighborhood, with larger values toward the Galactic center \cite{Haverkorn:2015,beck2016magnetic}. Similarly, in the Andromeda galaxy (M31), polarized radio observations reveal a well-ordered large-scale magnetic field with an average strength of about $6 \ \mu$G in the prominent ring located at galactocentric radii of $8$–$12$~kpc \cite{beck2016magnetic,Beck:2025}. Although Centaurus~A is morphologically distinct and hosts active galactic phenomena, polarimetric observations of its dust and gas indicate the presence of coherent magnetic-field structures on $kpc$ scales \cite{Lopez-Rodriguez:2021}.

In the present work, we combine these two ideas. We consider a complex SFDM halo endowed with a local $U(1)$ symmetry, such that the scalar field is minimally coupled to a gauge field $B_{\mu}$. We treat the gauge field as a perturbation and study the evolution of scalar and gauge perturbations in an expanding Universe. We show that the spatial structure of the gauge field inherits the same spherical Bessel and harmonic modes that characterize the multistate SFDM halo, and that the presence of the gauge field does not modify the underlying dark-matter density profile obtained in \cite{bernal2025natural}. In this way, the SFDM halo acts as a geometric framework that sources a large-scale magnetic field whose morphology is determined by the excited modes of the scalar field. In this way, we present a natural explanation for both phenomena, the VPOS observed in the Milky Way, Andromeda, and Centaurus A, and their magnetic fields.

For a start, we assume that the Universe is homogeneous and isotropic, and that structure formation arises from density perturbations that grow throughout cosmic history. These perturbations are directly related to fluctuations of the scalar field. In Section~\ref{sec:perturbed}, we present the equations governing the evolution of scalar-field perturbations in the presence of the gauge field $B_{\mu}$. In Section~\ref{sec:magnetic}, we find the spatial components of $B_{\mu}$, which subsequently allow us to compute the associated electric and magnetic fields. In Section~\ref{sec:background}, we describe the background solutions required to solve the perturbed system and determine the time evolution of the scalar-field perturbations. This temporal evolution modifies only the time dependence of the fluctuations, leaving their spatial distribution unchanged. We also show that the presence of the gauge field does not alter the scalar-field density distribution previously obtained in~\cite{bernal2025natural,matos2022quantum}, as discussed in Subsection~\ref{subsec:spatialdis}. In Subsection~\ref{sec:potential}, we compute the corresponding gravitational potential. In Section~\ref{sec:results}, we present the results of the resulting magnetic fields for the Milky Way, Andromeda, and Centaurus A. Finally, our main conclusions are summarized in Section~\ref{sec:conclusions}.

Throughout the paper, we work in units where $c=\hbar=1$, and we adopt the metric signature $(-,+,+,+)$. Greek indices run over spacetime coordinates, while Latin indices label spatial coordinates only.

\section{Perturbed equations}
\label{sec:perturbed}

We start with the Lagrangian density:
\begin{equation}
\label{eq:lagrangiandensity1}
    \mathcal{L}=-(\nabla_{\mu}\Phi+iqB_{\mu}\Phi)(\nabla^{\mu}\Phi^{*}-iqB^{\mu}\Phi^{*})-V(\Phi\Phi^{*})-\frac{1}{4}B_{\mu\nu}B^{\mu\nu}-\frac{1}{4}F_{\mu\nu}F^{\mu\nu}-\frac{\delta}{2}B_{\mu\nu}F^{\mu\nu},
\end{equation}
where $F_{\mu\nu}=\nabla_{\mu}A_{\nu}-\nabla_{\nu}A_{\mu}$ is the electromagnetic gauge field of the standard model, $\delta$ is a kinetic coupling constant, and $B_{\mu\nu}=\nabla_{\mu}B_{\nu}-\nabla_{\nu}B_{\mu}$, where $B_{\mu}$ is the gauge field of the SFDM. 

This Lagrangian describes a complex charged scalar field coupled to the Maxwell field. The gauge field of the scalar field only interacts with the gauge field of the standard model, the ``photon". Then it can be found that: 
\begin{eqnarray}
    \nabla_{\mu}F^{\mu\nu}&=&-\delta\nabla_{\mu}B^{\mu\nu},\nonumber\\
    \nabla_{\mu}B^{\mu\nu}+\delta\nabla_{\mu}F^{\mu\nu}&=&J^{\nu}_{\Phi},
\end{eqnarray}
where the conserved 4-current $J_\mu$ is defined as: 
\begin{equation}
    J_{\mu}=\frac{iq}{2m^2}\left[\Phi(\nabla_{\mu}-iqB_{\mu})\Phi^{*}-\Phi^{*}(\nabla_{\mu}+iqB_{\mu})\Phi\right].
\end{equation}
The above equations tell us that the electromagnetic field is generated by this dark photon, but the strength of this electromagnetic field depends on the value of the mixing term $\delta$. In the above equations, we are assuming that the dark photon and the photon of the standard model are different. However, it is also possible that the gauge field associated with the scalar field coincides with that of the standard model. In such a case, the scalar field should carry a very small charge in order to remain consistent with observational constraints while providing a viable explanation for the origin of galactic magnetic fields. Following this idea, the mixing term is not necessary, and we consider the following Lagrangian:  
\begin{equation}
\label{eq:lagrangiandensity2}
\mathcal{L}_1=-(\nabla_{\mu}\Phi+iqB_{\mu}\Phi)(\nabla^{\mu}\Phi^{*}-iqB^{\mu}\Phi^{*})-V(\Phi\Phi^{*})-\frac{1}{4}B_{\mu\nu}B^{\mu\nu},
\end{equation}
with the scalar field potential \citep[see][]{bernal2025natural}:
\begin{equation}
\label{eq:potential}
    V(\Phi\Phi^{*})=\frac{\lambda}{2}\left(\Phi\Phi^{*}-\frac{m^2}{\lambda}+\frac{1}{2}T^2\right)^2 ,
\end{equation}
where $m$ is the mass of the scalar field, $\lambda$ the self-interaction parameter between SF particles, and $T$ the temperature of the thermal bath.
We denote $\Phi\Phi^{*}:=n$ as the scalar field number density.

Then, the energy-momentum tensor is given by: 
\begin{eqnarray}
\label{eq:energytensor}
    T_{\mu\nu}&=&-g_{\mu\nu}\left[g^{\alpha\beta}(\nabla_{\alpha}\Phi+iqB_{\alpha}\Phi)(\nabla_{\beta}\Phi^{*}-iqB_{\beta}\Phi^{*})+V(\Phi\Phi^{*})\right]\nonumber\\
    &+&\left(\nabla_{\mu}\Phi+iqB_{\mu}\Phi\right)\left(\nabla_{\nu}\Phi^{*}-iqB_{\nu}\Phi^{*}\right)+\left(\nabla_{\mu}\Phi^{*}-iqB_{\mu}\Phi^{*}\right)\left(\nabla_{\nu}\Phi+iqB_{\nu}\Phi\right)\nonumber\\
    &-&\frac{1}{4}g_{\mu\nu}B_{\alpha\beta}B^{\alpha\beta}+B^{\alpha}_{\mu}B_{\alpha\nu},
\end{eqnarray}
while the field equation for the scalar field is given by:
\begin{equation}
\label{eq:SF}
    \Box_E\Phi-\frac{dV}{d\Phi^{*}}=0,
\end{equation}
where 
\begin{eqnarray}
\label{eq:dalambert}
    \Box_E&=&(\nabla^{\mu}+iqB^{\mu})(\nabla_{\mu}+iqB_{\mu}), \\
\label{eq:maxwell}
    \nabla_{\nu}B^{\nu\mu}&=&J^{\mu}.
\end{eqnarray}

To obtain the perturbed equations, we consider the following perturbed metric:
\begin{equation}
\label{eq:pertubedmetric}
    ds^2=a^2(\eta)\left[ -(1+2\psi)d\eta^2+(1-2\phi)\delta_{ij}dx^idx^j \right].
\end{equation}
Here, $\eta$ is the conformal time, and $d\eta=dt/a$, where $t$ is the cosmic time and $a$ the scale factor. Taking $\Phi=\Phi_0(\eta)+\delta\Phi(\eta,x^i)$ in equation \eqref{eq:SF}, and using the perturbed metric \eqref{eq:pertubedmetric}, we obtain the Klein-Gordon equation for the background given by:
\begin{equation}
\label{eq:KG1}
    \ddot{\Phi}_0+2\mathcal{H}\dot{\Phi}_0+a^2\frac{dV}{d\Phi^{*}}\bigg|_{\Phi_0^*}=0,
\end{equation}
where $\mathcal{H}=(1/a)(da/d\eta)=aH$, with $H=(1/a)(da/dt)$ the usual Hubble parameter, and $V$ the potential for the scalar field after symmetry breaking~\citep[see][]{bernal2025natural}:
\begin{equation}
\label{eq:potential_after}
    V=\left(m^2-\frac{\lambda}{2}T^2\right)\Phi\Phi^*+\frac{\lambda}{2}(\Phi\Phi^*)^2,
\end{equation}
where we can define $m_\Phi := m^2-(\lambda/2)T^2$ as the effective mass of the scalar field.

The equation for the perturbation $\delta\Phi$ is then: 
\begin{equation}
\label{KG:per}
  \nabla^2\delta\Phi  -{\delta\ddot{\Phi}}-2\mathcal{H}\delta{\dot{\Phi}}+{4}\dot{\phi}\dot{\Phi}_0-2iqB_0\dot{\Phi}_0-a^2\left(V_{,\Phi_0\Phi_0^{*}}\delta\Phi+V_{,\Phi_0^*\Phi_0^*}\delta\Phi^*+2V_{,\Phi_0^*}\phi\right)=0,
\end{equation}
where we have assumed the Lorenz condition $\nabla_{\mu}B^{\mu}=0$. After a Mandelung transformation for the scalar field of the form:
\begin{equation}
\Phi(\textbf{x},t)=R(\textbf{x},t)e^{i\theta(\textbf{x},t)},
\end{equation}
and taking again $\Phi=\Phi_{0}(\eta)+\delta\Phi(\eta,x^i)$, we find:
\begin{equation}
    \Phi=\Phi_0+\left.\frac{\partial\Phi}{\partial R}\right|_0(R-R_0)+\left.\frac{\partial\Phi}{\partial\theta}\right|_0(\theta-\theta_0)+...,
\end{equation}
where $|_0=|_{R_0,\theta_0}$ and $\Phi_0=R_0(\eta)e^{i\theta_0(\eta)}$. For the scalar field perturbation we have:
\begin{equation}
\label{eq:deltaphi}
    \delta\Phi=\left[\delta R(\textbf{x},\eta)+iR_0\delta\theta(\textbf{x},\eta)\right]e^{i\theta_0},
\end{equation}
Then, if we replace $\Phi_0=R_0e^{i\theta}$ in equation \eqref{eq:KG1}, we obtain for the real and imaginary parts, respectively, the following: 
\begin{eqnarray}
\label{eq:backgroundR0}
    \ddot{R}_0-{R}_0\dot{\theta}_0^2+2\mathcal{H}\dot{R}_0+a^2\frac{dV}{d\Phi^*}&=&0,\\
    \label{eq:backgroundtheta0}
     \ddot{\theta}_0+2\mathcal{H}\dot{\theta}_0+2\dot{\theta}_0\frac{\dot{R}_0}{R_0}&=&0,\\
     \label{eq:backgndtheta0}
     \frac{d}{d\eta}\left(\dot{\theta}_0R_0^2a^2\right)&=&0,
\end{eqnarray}
where equation \eqref{eq:backgndtheta0} is obtained from \eqref{eq:backgroundtheta0}.
Then, if we replace \eqref{eq:deltaphi} in \eqref{KG:per}, and if we consider the Lorenz condition, we obtain for the real part of \eqref{KG:per} the following equation: 
\begin{eqnarray}
    &&\nabla^2\delta R-\delta\ddot{R}-2\mathcal{H}\delta\dot{R}+\left[\dot{\theta}_0^2-\left(m^2-\frac{\lambda T^2}{2}\right)a^2-3\lambda R_0^2 a^2\right]\delta R\nonumber\\
    &+&\left(R_0\ddot{\theta}_0+2\dot{R_0}\dot{\theta}_0+2\mathcal{H}R_0\dot{\theta}_0\right)\delta\theta+2R_0\dot{\theta}_0\delta\dot{\theta}+4\dot{R}_0\dot{\phi}+2qB_0R_0\dot{\theta}_0=0.
\end{eqnarray}
Taking now $\delta\dot{\theta}\approx 0$ and using equation \eqref{eq:backgroundtheta0} which implies that the factor that multiplies $\delta\theta$ is zero, we finally obtain: 
\begin{equation}
\label{eq:deltaR2}
\nabla^2\delta R-\delta\ddot{R}-2\mathcal{H}\delta\dot{R}+\left[\dot{\theta}_0^2-\left(m^2-\frac{\lambda T^2}{2}\right)a^2-3\lambda R_0^2 a^2\right]\delta R+4\dot{R}_0\dot{\phi}+2qB_0R_0\dot{\theta}_0=0.
\end{equation}
For the imaginary part of \textcolor{blue}{\eqref{KG:per}}, we have: 
\begin{equation}
    \delta \dot{R}+\left(\frac{\ddot{\theta}}{2\dot{\theta}}+\mathcal{H}\right)\delta R -2\dot{\phi}R_0-qB_0\frac{\dot{R}_0}{R_0}=0,
\end{equation}
and using the background equation (\ref{eq:backgroundtheta0}) for $\theta$, we have:
\begin{equation}
\label{eq:deltaR}
    \delta\dot{R}-\frac{\dot{R}_0}{R_0}\delta R-2\dot{\phi}R_0-qB_0\frac{\dot{R}_0}{\dot{\theta_0}}=0,
\end{equation}
and solving for $B_0$: 
\begin{equation}
\label{eq:B0}
    B_0=\frac{\dot{\theta}_0}{q}\left(-\frac{\delta\dot{R}}{\dot{R}_0}+\frac{\delta R}{R_0}+2\dot{\phi}\frac{R_0}{\dot{R}_0}\right) .
\end{equation}
With the last equation, we can see that the evolution of the time component of the gauge field $B_\mu$ is related to the evolution of the perturbation and the background scalar field, as in a previous work \cite{bernal2025natural}. If we now consider $\dot{\phi}\approx 0$, and we replace \eqref{eq:B0} in \eqref{eq:deltaR2}, then we obtain a second differential partial equation for $\delta R$:
    \begin{equation}
    \label{eq:deltaR3}
    \nabla^2\delta R-\delta\ddot{R}-2\left(\mathcal{H}+\frac{R_0\dot{\theta}_0^2}{\dot{R}_0}\right) \dot{\delta R}
    +\left[3\dot{\theta}_0^2+a^2\left(-m^2+\frac{\lambda}{2}T^2\right)-3\lambda R_0^2a^2\right]\delta R=0.
\end{equation}
To solve it, we propose the following ansatz: 
\begin{equation}
\label{eq:dR}
    \delta R=\sum_{j,k}A_{jk}L_k(r)Y_k^j(\theta,\varphi) T_R(\eta),
\end{equation}
where $Y_{k}^j(\theta,\varphi)$ are the harmonic spherical polynomials, $L_k(r)$ is a function that depends only on the radial coordinate $r$, and $T_R(\eta)$ is a function that depends on the conformal time. Then we obtain from \eqref{eq:deltaR3} the following differential equation for $L_k$:
\begin{equation}
\label{eq:diffeqLk}
    \frac{\partial}{\partial r}\left(r^2\frac{\partial L_k}{\partial r}\right)+\left[l^2r^2-k(k+1)\right]L_k=0,
\end{equation}
where $l$ is a constant that can in principle take any value. In particular, in \cite{bernal2025natural} it was found that the value of $l$ is of the order of ${kpc^{-1}}$ for the Milky Way, Andromeda, Centaurus A, and six other Milky Way-like galaxies.
If we now make the change of variables $x=lr$, we obtain from \eqref{eq:diffeqLk}:
\begin{equation}
    x^2\frac{d^2L_k}{dx^2}+2x\frac{dL_k}{dx}+\left[x^2-k(k+1)\right]L_k=0,
\end{equation}
whose solution are the spherical Bessel functions $j_k(x)=L_k(x)$, which are related with the Bessel functions by means of $j_k(x)=\sqrt{\frac{\pi}{2x}}J_{k+1/2}(x)$, and whose first values are: 
\begin{eqnarray}
    j_0(x)&=&\sqrt{\frac{\pi}{2x}}J_{1/2}(x)=\frac{\sin x}{x},\\
    j_1(x)&=&\sqrt{\frac{\pi}{2x}}J_{3/2}(x)=\frac{\sin x}{x^2}-\frac{\cos x}{x}.
\end{eqnarray}
Therefore, we have found that: 
\begin{equation}
    \delta R=\sum_{k,m}A_{km}j_k(lr)Y_k^m(\theta,\varphi)T_R.
\end{equation}
But from equation (\ref{eq:deltaphi}) we know that $\delta R$ is a real function, then: 
\begin{equation}
\label{deltaR}
    \delta R=\sum_k A_{k0}j_k(lr)Y_k^0(\theta,\varphi)T_R,
\end{equation}
where $T_R$ satisfies the following differential equation:
\begin{equation}
    \label{eq:TReta}\ddot{T_R}+2\left(\frac{R_0\dot{\theta}_0^2}{\dot{R}_0}+\mathcal{H}\right)\dot{T}_R+\left(\omega^2+3\lambda R_0^2a^2-3\dot{\theta}_0^2\right)T_R=0,
\end{equation} 
where here $\omega^2=l^2+\frac{\lambda}{2}(T_c^2-T^2)a^2=l^2+(m^2-\frac{\lambda T^2}{2})a^2=l^2+m_{\Phi}^2a^2$.\\

The density of the fluctuation that collapses to form structure is $\delta\Phi\delta\Phi^*=\delta R^2+R_0^2\delta\theta^2\approx \delta R^2$, then the spatial distribution of the perturbation is still described by spherical Bessel functions multiplied by spherical harmonics as in \cite{bernal2025natural}, despite the presence of a gauge field. Also, this solution shows that there are different configurations for the halo. The appearance of spherical Bessel functions in the radial part and spherical harmonics in the angular dependence has precisely the same structure found for the multistate SFDM halos in \cite{bernal2025natural}. This means that the scalar perturbations that seed structure formation share the same ``atomic'' mode structure, and as we will see below, the gauge-field components also inherit this structure.

\begin{figure}[h]
    \centering
    \includegraphics[width=0.45\textwidth]{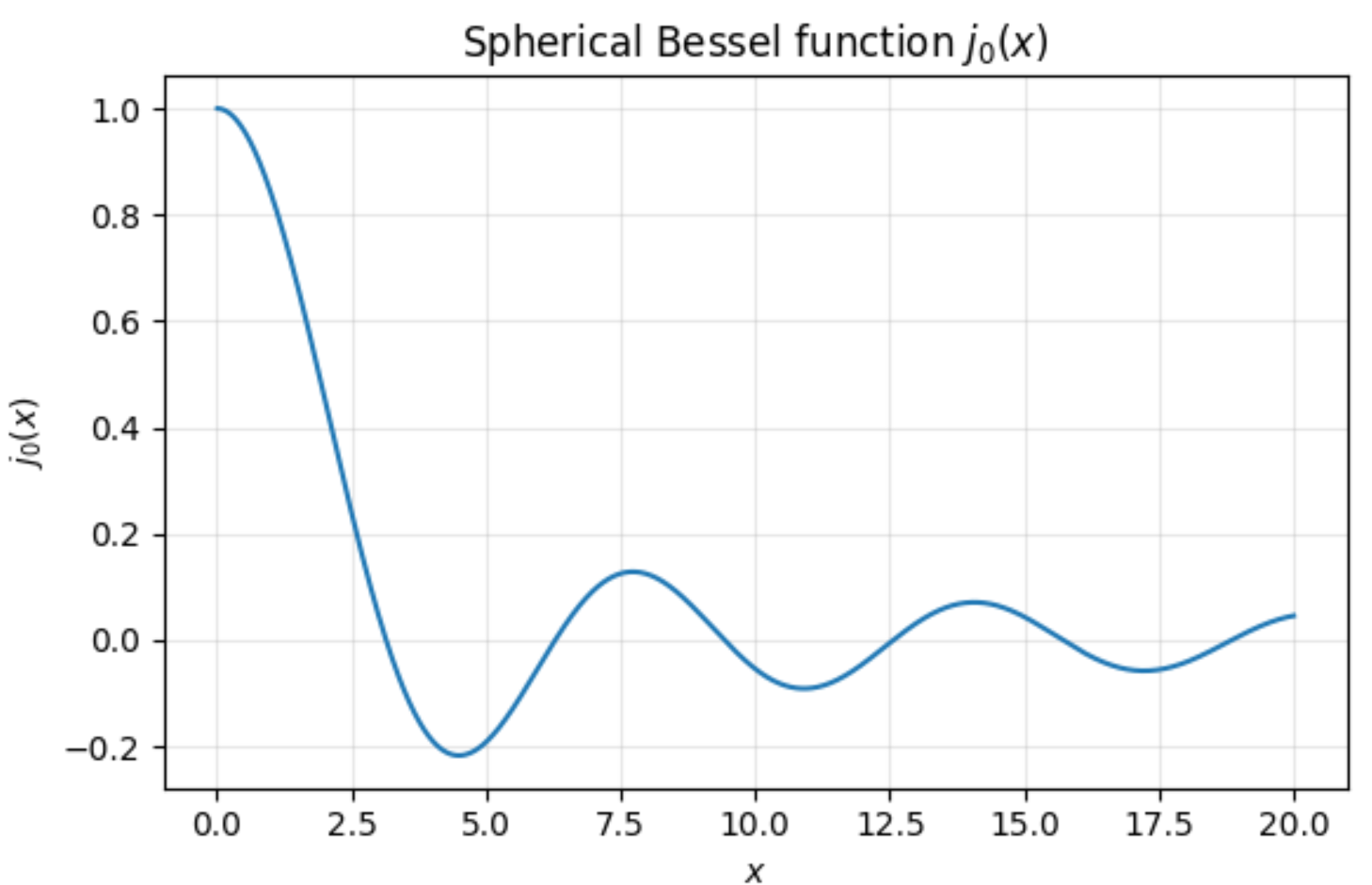}
    \includegraphics[width=0.45\textwidth]{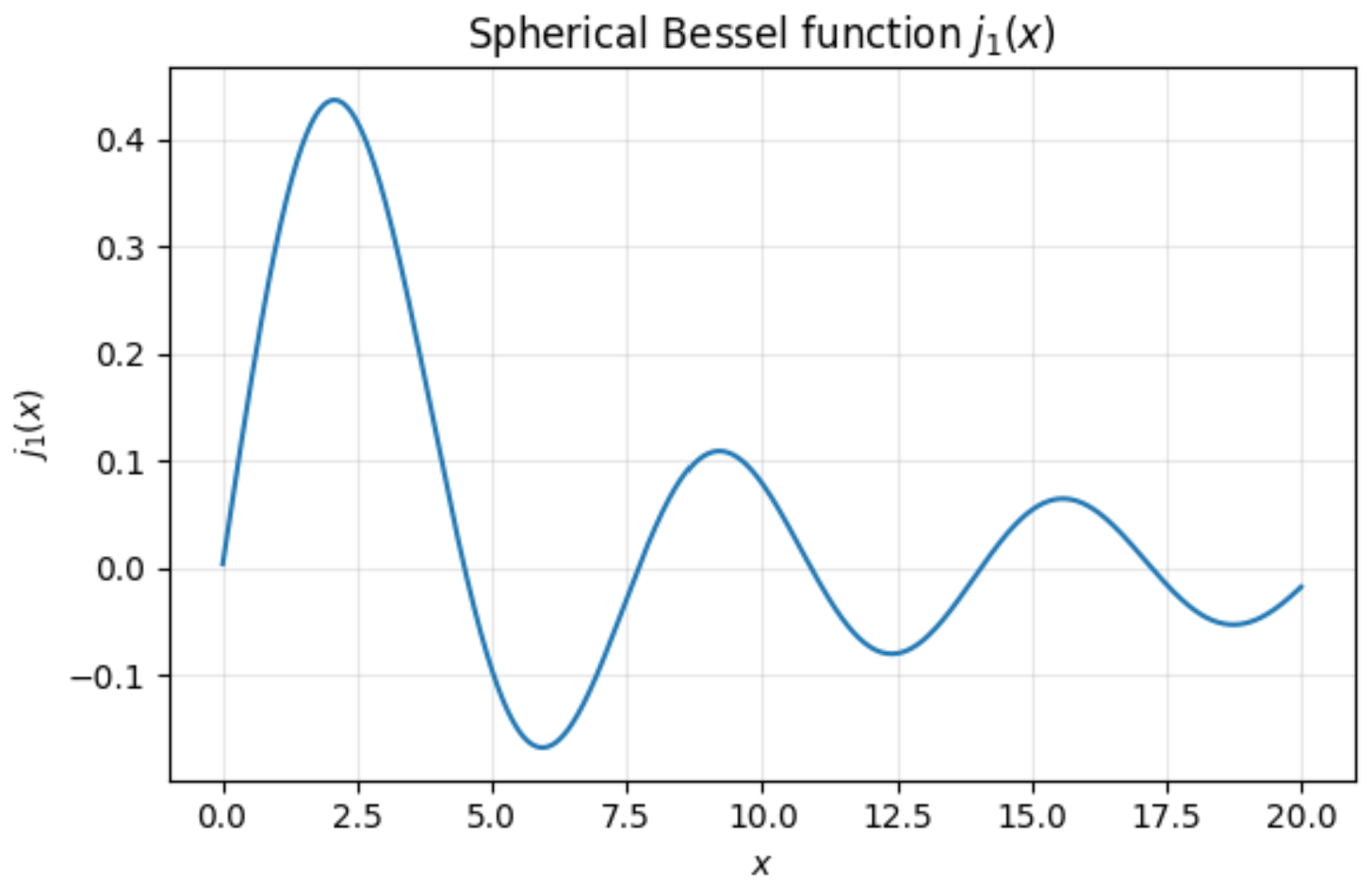}
    \caption{Radial dependence of the first two spherical Bessel functions, $j_0(x)$ (left) and $j_1(x)$ (right). The function $j_0$ corresponds to the spherically symmetric ground state, while $j_1$ vanishes at the origin and describes the first excited, dipolar-like mode that plays a central role in the multistate SFDM halo and the associated gauge-field configurations.}
    \label{fig:j0j1}
\end{figure}

To illustrate the behavior of the lowest modes, in Fig.~\ref{fig:j0j1} we show the radial dependence of $j_0(x)$ and $j_1(x)$, which correspond to the ground state and first excited state, respectively. The ground state is finite at the origin and decays with oscillations, while the first excited state vanishes at the origin and exhibits a nodal structure consistent with a dipolar configuration.

Furthermore, using (\ref{eq:B0}) and (\ref{eq:dR}) we find that: 
\begin{equation}
\label{eq:B03}
    B_0=\frac{\dot{\theta}_0}{q}\sum_kA_{k0}j_k(lr)Y_k^0(\theta,\varphi)\left(\frac{T_R}{R_0}-\frac{\dot{T}_R}{\dot{R}_0}\right)=\sum_kA_{k0}T_0(\eta)j_k(lr)Y_k^0(\theta,\varphi),
\end{equation}
where we have defined
\begin{equation}
\label{eq:T0}
    T_0(\eta)=\frac{\dot{\theta}_0}{q}\left(\frac{T_R}{R_0}-\frac{\dot{T}_R}{\dot{R}_0}\right).
\end{equation}
Therefore, the gauge field can also be written as a product of a Bessel function that depends on $l$, multiplied by spherical harmonics and a function of time.

In order to find $T_R$ we now need to solve equation \eqref{eq:TReta}, but first we need to solve for the background. To do this, we can rewrite the background equations in terms of the variable $N=\ln(a)$; this will be done in Section \ref{sec:background}.  In terms of $N$, equation \eqref{eq:TReta} can be rewritten as: 
\begin{equation}
\label{eq:TR}
    T_R''+\left(3+\frac{\dot{H}}{aH^2}+2\frac{R_0\theta_0'^2}{R_0'}\right)T_R'+\left(\frac{\omega^2}{a^2H^2}+\frac{3\lambda R_0^2}{H^2}-3\theta_0'^2\right)T_R=0,
\end{equation}
where the prime $'$ denotes derivative with respect to $N=\ln(a)$.

\section{Magnetic and electric fields}
\label{sec:magnetic}

From equation \eqref{eq:maxwell}
 and using the Lorenz, condition $\nabla_{\mu}B^{\mu}=0$, we obtain: 
\begin{eqnarray}
\label{eq:B02}
    \ddot{B}_0+2\left(\dot{\mathcal{H}}B_0+\mathcal{H}\dot{B}_0\right)-\nabla^2B_0&=&-\frac{a^2}{m_{\Phi}^2}\left(qR_0^2\dot{\theta}_0+q^2R_0^2B_0+2qR_0\dot{\theta}_0\delta R\right) , \\
\label{eq:Bi}
    \partial_j\partial_jB_i-\ddot{B}_i-2\mathcal{H}\partial_iB_0 &=&2a^2q^2B_iR_0^2.
\end{eqnarray}
In the previous section, we showed that $B_0$ can be written in terms of $\delta R$ (equation \eqref{eq:B0}), but at the same time $B_0$ has to satisfy (\ref{eq:B02}). In this section, we are going to assume that $B_0=f_0(r)\sum_{p,q}
Y_{p}^qM_{pq}(\theta,\varphi) \mathcal{T}_{0}(\eta)$,  with this we find: 
\begin{eqnarray}
\label{eq:f0}
   \sum_{p,q}M_{pq}Y_{p}^q(\theta,\varphi)\left[ \frac{1}{r^2}\frac{d}{d r}\left(r^2\frac{d f_{0}}{d r}\right)+\left(\Omega_0^2-\frac{p(p+1)}{r^2}\right)f_0\right]&-&\sum_kA_{k0}\frac{2qa^2R_0\dot{\theta}_0 T_R}{m_{\Phi}^2\mathcal{T}_{0}}j_k(lr)Y_{k}^0(\theta,\varphi) \nonumber \\ 
   &=&\frac{a^2q}{m_{\Phi}^2}\left(\frac{R_0^2\dot{\theta_0}}{\mathcal{T}_{0}}\right),
\end{eqnarray}
where $\frac{\ddot{\mathcal{T}}_{0}}{\mathcal{T}_{0}}+2(\dot{\mathcal{{H}}}+\mathcal{H}\frac{\dot{\mathcal{T}}_{0}}{\mathcal{T}_{0}})+\frac{a^2q^2R_0^2}{m_{\Phi}^2}=-\Omega_{0}^2$, with $\Omega_0=$ constant. From \eqref{eq:backgndtheta0} we see that the term $a^2R_0^2\dot{\theta}_0=$ constant, and there is only a solution if $p=0,q=0$ and $k=0$; then we obtain: 
\begin{equation}
    \frac{1}{r^2}\frac{d}{dr}\left(r^2\frac{df_0}{dr}\right)+\Omega_0^2f_0=\frac{b_0}{\mathcal{T}_0}+c_0\frac{T_R}{\mathcal{T}_0}j_0(lr),
\end{equation}
where $b_0=\sqrt{4\pi}a^2q\dot{\theta}_0R_0^2/(m_{\Phi}^2M_{00})$ and $c_0=2qa^2R_0\dot{\theta}_0A_{00}/(M_{00}m_{\Phi}^2)$. Then, the solution to the homogeneous equation is: 
\begin{equation}
    f_{0}(r)=Cj_0(\Omega_{0}r),
\end{equation}
where $C$ is a constant that has to be determined. The final solution has the form: 
\begin{equation}
    f_0(r)=Cj_0(\Omega_0r)+\frac{c_0T_R}{(\Omega_0^2-l ^2)\mathcal{T}_0}j_0(lr)+\frac{b_0}{\Omega_0^2\mathcal{T}_0},
\end{equation}
and since $B_0=f_0(r)Y^0_0(\theta,\varphi)\mathcal{T}_0$, we find: 
\begin{equation}
\label{b02}
    B_0=Cj_0(\Omega_0 r)Y^0_0(\theta,\varphi)\mathcal{T}_0+\frac{c_0T_RY_0^0(\theta,\varphi)}{(\Omega_0^2-l^2)}j_0(lr)+\frac{b_0}{\Omega_0^2}.
\end{equation}
But this solution has to match with equation (\ref{eq:B03}) that we rewrite it here in order to compare it with \eqref{b02}:
\begin{equation}
 B_0=\sum_kA_{k0}\frac{\dot{\theta}_0}{q}j_k(lr)Y_k^m(\theta,\varphi)\left(\frac{T_R}{R_0}-\frac{\dot{T}_R}{\dot{R}_0}\right)=\sum_kA_{k0}T_0(\eta)j_k(lr)Y_k^m(\theta,\varphi),\nonumber
\end{equation}
thus $C=0$, $(b_0/\Omega_0^2)\sim 0$, and 
\begin{equation}
\label{eq:c0TR}
    \frac{c_0T_R}{\Omega_0^2-l^2}=\frac{\dot{\theta}_0}{q}\left(\frac{T_R}{R_0}-\frac{\dot{T}_R}{\dot{R}_0}\right)=A_{00}T_0= \text{constant}.
\end{equation}
Then from \eqref{eq:c0TR}, we can find that: 
\begin{equation}
\label{eq:dTR}
    \dot{T}_R=qT_0\frac{\dot{R_0}}{\dot{\theta}_0}-\frac{\dot{R}_0}{R_0}T_R.
\end{equation}
Therefore $T_R$ has to satisfy \eqref{eq:TReta} and \eqref{eq:dTR}.
The most important result of this section is that 
\begin{equation}
    B_0=\frac{j_0(lr)A_{00}T_0}{2\sqrt{\pi}},
\end{equation}
where $T_0$ (equation \eqref{eq:c0TR}) is a constant, and this solution is consistent with equation \eqref{eq:B03}.

Now, in order to find $B_x$, we again assume that: 
\begin{equation}
   B_x= f_{x}(r)T_{x}(\eta)\sum_{u,v}N_{uv}Y_u^v(\theta,\varphi).
\end{equation}
If we replace this ansatz in \eqref{eq:Bi}, we obtain the following partial differential equation for $f_{x}(r)$:
\begin{equation}
\sum_{u,v}N_{uv}Y_{u}^v(\theta,\varphi)\left[\frac{1}{r^2}\frac{d}{d r}\left(r^2\frac{d f_x}{d r}\right)+\left(\Omega_x^2-\frac{u(u+1)}{r^2}\right)f_x\right]=\frac{\mathcal{H}T_0A_{00}\sin\theta\cos\varphi}{\sqrt{\pi}T_x}\frac{d(j_0(lr))}{d r},
\end{equation}
where
\begin{equation}
    \Omega_{x}^2=-\frac{\ddot{T}_{x}}{T_{x}}-2a^2q^2 R_0^2,
\end{equation}
and since $\sqrt{\frac{2\pi}{3}}\left(Y_1^{-1}(\theta,\varphi)-Y_1^1(\theta,\varphi)\right)=\sin\theta\cos\varphi$, then the only terms $N_{uv}$ different from zero are $N_{1-1},N_{11}$, and we have:
\begin{eqnarray}
   (N_{1-1}Y^{-1}_1(\theta,\varphi)+N_{11}Y_1^1(\theta,\varphi))&&\left[\frac{1}{r^2}\frac{d}{d r}\left(r^2\frac{d f_x}{d r}\right)+\left(\Omega_x^2-\frac{2}{r^2}\right)f_x\right]= \\
   &&\frac{\mathcal{H}T_0A_{00}}{\sqrt{\pi}T_x}\sqrt{\frac{2\pi}{3}}\frac{d(j_0(lr))}{d r}(Y_1^{-1}(\theta,\varphi)-Y_1^1(\theta,\varphi)). \nonumber
\end{eqnarray}
Then from these expressions we have: 
\begin{equation}
\label{eq:N11}
    N_{1-1}=-N_{11}=\frac{\mathcal{H}T_0A_{00}}{T_x}\sqrt{\frac{2}{3}}.
\end{equation}
The differential equation for $f_x$ becomes:
\begin{equation}
    \frac{1}{r^2}\frac{d}{dr}\left(r^2\frac{d f_x}{dr}\right)+\left(\Omega_x^2-\frac{2}.{r^2}\right)f_x=-lj_1(lr)
\end{equation}
The solutions for the homogeneous and particular equations are:
\begin{eqnarray}
    f_{xh}&=&c_xj_1(\Omega_xr) , \\
    f_{xp}&=&-\frac{l}{\Omega_x^2-l^2}j_1(lr) ,
\end{eqnarray}
and the total solution is given by
\begin{equation}
    f_x=f_{xh}+f_{xp}=c_xj_1(\Omega_xr)-\frac{l}{\Omega_x^2-l^2}j_1(lr),
\end{equation}
where $c_x$ is a constant that has to be determined.

Then, for $B_x$ we finally have: 
\begin{equation}
    B_x=f_xT_x\left[N_{1-1}Y_1^{-1}(\theta,\varphi)+N_{11}Y^1_1(\theta,\varphi)\right]=\left(c_xj_1(\Omega_xr)-\frac{l}{\Omega_x^2-l^2}j_1(lr)\right)\frac{\mathcal{H}T_0A_{00}\sin\theta\cos\varphi}{\sqrt{\pi}}. 
\end{equation}

For $B_y=f_y(r)T_y(\eta)\sum_{s,w}L_{sw}Y_s^w(\theta,\varphi)$, and similarly we find:
\begin{equation}
\sum_{s,w}L_{sw}Y_{s}^w(\theta,\varphi)\left[\frac{1}{r^2}\frac{d}{d r}\left(r^2\frac{d f_x}{d r}\right)+\left(\Omega_x^2-\frac{s(s+1)}{r^2}\right)f_x\right]=\frac{\mathcal{H}T_0A_{00}\sin\theta\sin\varphi}{\sqrt{\pi}T_x}\frac{d(j_0(lr))}{d r},
\end{equation}
and the only terms that are different from zero are: 
\begin{equation}
\label{eq:L11}
    L_{11}=L_{1-1}=i\sqrt{\frac{2}{3}}A_{00}\frac{\mathcal{H}T_0}{T_y},
\end{equation}
so the differential equation for $f_y$ becomes:
\begin{equation}
    \frac{1}{r^2}\frac{d}{dr}\left(r^2\frac{d f_y}{dr}\right)+\left(\Omega_y^2-\frac{2}{r^2}\right)f_y=-lj_1(lr) .
\end{equation}
Then 
\begin{equation}
    B_y=f_yT_y\left[L_{1-1}Y_1^{-1}(\theta,\varphi)+L_{11}Y^1_1(\theta,\varphi)\right]=\left(c_yj_1(\Omega_yr)-\frac{l}{\Omega_y^2-l^2}j_1(lr)\right)\frac{\mathcal{H}T_0A_{00}\sin\theta\sin\varphi}{\sqrt{\pi}}. 
\end{equation}

Finally, for $B_z=f_z(r)T_z(\eta)\sum_{a,b}Q_{ab}Y_a^{b}(\theta,\varphi)$ we find that the only term different from zero is:
\begin{equation}
\label{eq:Q10}
    Q_{10}=\sqrt{\frac{4}{3}}\mathcal{H}\frac{T_0}{T_z}.
\end{equation}
and in the same way, we find: 
\begin{equation}
    B_{z}=\left[c_zj_1(\Omega_{z}r)-\frac{l}{\Omega_z^2-l^2}j_1(lr)\right]\frac{\mathcal{H}A_{00}T_0\cos\theta}{\sqrt{\pi}},
\end{equation}
where $\Omega_z=-\frac{\ddot{T}_z}{T_z}-2a^2q^2R_0^2$.\\

In summary, we have found that:
\begin{eqnarray}
B_0&=&\frac{1}{2\sqrt{\pi}}j_0(lr)A_{00}T_0=\frac{1}{2\sqrt{\pi}}\frac{\sin(lr)}{lr}A_{00}T_0 ,
\nonumber\\
   B_{x}&=&\left[c_xj_1(\Omega_xr)-\frac{l}{\Omega_x^2-l^2}j_1(lr)\right]\frac{\mathcal{H}A_{00}T_0\sin\theta\cos\varphi}{\sqrt{\pi}},\nonumber\\
    B_{y}&=&\left[c_yj_1(\Omega_{y }r)-\frac{l}{\Omega_y^2-l^2}j_1(lr)\right]\frac{\mathcal{H}A_{00}T_0\sin\theta\sin\varphi}{\sqrt{\pi}},\nonumber\\
    B_{z}&=&\left[c_zj_1(\Omega_{z}r)-\frac{l}{\Omega_z^2-l^2}j_1(lr)\right]\frac{\mathcal{H}A_{00}T_0\cos\theta}{\sqrt{\pi}}.
\end{eqnarray}
Also, from equations \eqref{eq:N11}, \eqref{eq:L11}, and \eqref{eq:Q10}, we find that $T_x, T_y, T_z \propto \mathcal{H}$. 
Finally, we can compute the components of the electric and magnetic fields from the following relations: 
\begin{eqnarray}
    \mathcal{E}^{\mu}=u_{\nu}F^{\mu\nu},\hspace{0.5cm} \mathcal{B^{\mu}}=\frac{1}{2}\epsilon^{\mu\nu\alpha\beta}u_vF_{\alpha\beta},
\end{eqnarray}
where $u_{\nu}$ is the co-moving 4-velocity.

The components of the electric field different from zero are: 
\begin{eqnarray}
    \mathcal{E}^x&=&\frac{A_{00}T_0\sin\theta\cos\varphi}{a^3 \sqrt{\pi}}\left[lj_1(lr)\left(\frac{\dot{\mathcal{H}}}{\Omega_x^2-l^2}-\frac{1}{2}\right)-\dot{\mathcal{H}}c_xj_1(\Omega_x r)\right],\\
    \mathcal{E}^{y}&=&\frac{1}{a^3}(\partial_yB_0-\partial_{\eta}B_y)\nonumber\\
    &=&\frac{A_{00}T_0\sin\theta\sin\varphi}{a^3 \sqrt{\pi}}\left[lj_1(lr)\left(\frac{\dot{\mathcal{H}}}{\Omega_y^2-l^2}-\frac{1}{2}\right)-c_y\dot{\mathcal{H}}j_1(\Omega_y r)\right],\\
    \mathcal{E}^{z}&=&\frac{1}{a^3}(\partial_{z}B_0-\partial_{ \eta}B_{z}) \nonumber\\
    &=&\frac{A_{00}T_0\cos\theta}{a^3 \sqrt{\pi}}\left[lj_1(lr)\left(\frac{\dot{\mathcal{H}}}{\Omega_z^2-l^2}-\frac{1}{2}\right)-c_z\dot{\mathcal{H}}j_1(\Omega_z r)\right].
\end{eqnarray}
While the components of the magnetic field are: 
\begin{equation}
    \mathcal{B}^x=\frac{1}{a^3}(\partial_yB_z-\partial_zB_y),\hspace{0.1cm} \mathcal{B}^y=-\frac{1}{a^3}(\partial_xB_z-\partial_zB_x), \hspace{0.1cm} \mathcal{B}^z=\frac{1}{a^3}(\partial_xB_y-\partial_y B_x),
\end{equation}
Then, we obtain: 
\begin{align}
\label{eq:Bx}
    \mathcal{B}^x=\frac{\sin\theta\cos\theta\sin\varphi\ \mathcal{H}A_{00}T_0}{a^3\sqrt{\pi}}\left[c_z\Omega_zj_0(\Omega_zr)-c_y\Omega_yj_0(\Omega_yr)+\frac{3}{r}\left(c_yj_1(\Omega_yr)-c_zj_1(\Omega_z r)\right)\right. \nonumber\\
    \left.+\left(\frac{1}{\Omega_z^2-l^2}-\frac{1}{\Omega_y^2-l^2}\right)\left(3l\frac{j_1(lr)}{r}-l^2j_0(lr)\right)\right],
\end{align}
\begin{align}
\label{eq:By}
    \mathcal{B}^y=\frac{\sin\theta\cos\theta\cos\varphi\ \mathcal{H}A_{00}T_0}{a^3\sqrt{\pi}}\left[c_x\Omega_xj_0(\Omega_xr)-c_z\Omega_zj_0(\Omega_zr)+\frac{3}{r}\left(c_zj_1(\Omega_zr)-c_xj_1(\Omega_x r)\right)\right.\nonumber\\
     \left.+\left(\frac{1}{\Omega_x^2-l^2}-\frac{1}{\Omega_z^2-l^2}\right)\left(3l\frac{j_1(lr)}{r}-l^2j_0(lr)\right)\right],
\end{align}
\begin{align}
\label{eq:Bz}
     \mathcal{B}^z=\frac{\sin\theta^2\sin\varphi\cos\varphi\ \mathcal{H}A_{00}T_0}{a^3\sqrt{\pi}}\left[c_y\Omega_yj_0(\Omega_yr)-c_x\Omega_xj_0(\Omega_xr)+\frac{3}{r}\left(c_xj_1(\Omega_xr)-c_yj_1(\Omega_y r)\right)\right.\nonumber\\
     \left.+\left(\frac{1}{\Omega_y^2-l^2}-\frac{1}{\Omega_x^2-l^2}\right)\left(3l\frac{j_1(lr)}{r}-l^2j_0(lr)\right)\right].
\end{align}
 And the total magnetic field is given by:
\begin{equation}
\mathcal{B}=\mathcal{B}^x\hat{i}+\mathcal{B}^y\hat{j}+\mathcal{B}^z\hat{k}.
\end{equation}

The expressions above show that the magnetic field components are built from linear combinations of $j_0$ and $j_1$ evaluated at different effective scales. Thus, the topology of the magnetic field lines is directly related to the nodal structure of the ground and first excited states of the SFDM halo, providing a link between the quantum nature of dark matter and the morphology of galactic magnetic fields.

\section{Background solution}
\label{sec:background}

For the background, we will consider the presence of the complex scalar field $\Phi_0$, baryons, radiation, and a cosmological constant $\Lambda$. Then the Friedmann equation is given by: 
\begin{equation}
\label{eq:friedmann}
    H^2=\frac{\kappa^2}{3}\left(\rho_{\Phi}+\rho_b+\rho_r+\rho_{\Lambda}\right),
\end{equation}
where $\rho_b$ is the density of baryons, $\rho_r$ is the radiation density, $\rho_{\Lambda}$ is the cosmological constant density, and $\rho_{\Phi}$ is the density of the scalar field $\Phi_0$, which in terms of the conformal time is given by: 
\begin{equation}
    \label{eq:rhophi}
    \rho_{\Phi}=\frac{\dot{\Phi}_0\dot{\Phi}^{*}_0}{a^2} .
\end{equation}
Each component of the Universe satisfies its corresponding continuity equation: 
\begin{equation}
\label{eq:continuity}
    \dot{\rho_b}=-3\mathcal{H}\rho_b, \ \dot{\rho_r}=-4\mathcal{H}\rho_r, \ \dot{\rho_{\Lambda}}=0, \ \dot{\rho_\Phi}=-6\mathcal{H}\frac{\dot{\Phi}_0\dot{\Phi}_0^*}{a^2}.
\end{equation}
Let us now define the following variables: 
\begin{equation}
\label{def:variables}
    x:=\frac{\kappa}{\sqrt{3}}\frac{\dot{\Phi}_0}{aH}, \ y:=\frac{\kappa}{\sqrt{3}}\frac{\dot{\Phi}_0^*}{aH}, \ u:=\frac{\kappa}{\sqrt{3}}\frac{m\Phi_0}{H}, \ 
    v:=\frac{\kappa}{\sqrt{3}}\frac{m\Phi_0^*}{H}, \
    b:=\frac{\kappa}{\sqrt{3}}\frac{\rho_b}{H}, \
    \Delta:=\frac{\kappa}{\sqrt{3}}\frac{\rho_{\Lambda}}{H}, \
    s:=\frac{m}{H}.
\end{equation}
If we derive equation \eqref{eq:friedmann} with respect to the conformal time $\eta$, and use the continuity equations \eqref{eq:continuity}, we obtain: 
\begin{equation}
\label{eq:Pi}
    \frac{3}{2}\Pi=-\frac{\dot{H}}{H^2}=3axy+\frac{3}{2}ab^2+2ar^2.
\end{equation}

From the Klein-Gordon equation \eqref{eq:KG1}, and from \eqref{eq:continuity}, we get the following system:  
\begin{eqnarray}
\label{ec:system}
        x'&=&-3x+\frac{3}{2}\frac{\Pi}{a}x-su,\hspace{0.5cm}
        y'=-3y+\frac{3}{2}\frac{\Pi}{a}y-sv,\nonumber\\
        u'&=&sx+\frac{3}{2}\frac{\Pi}{a}u,\hspace{1.7cm} v'=sy+\frac{3}{2}\frac{\Pi}{a}v,\nonumber\\
        b'&=&\frac{3}{2}\left(\frac{\Pi}{a}-1\right)b,\hspace{1.3cm}
        r'=\frac{3}{2}\frac{\Pi}{a}r-2r,\nonumber\\
        \Delta'&=&\frac{3}{2}\frac{\Pi}{a},\Delta\hspace{2.5cm}
        s'=\frac{3}{2}\frac{\Pi}{a}s,
\end{eqnarray}
where as before the prime $'$ denotes derivative with respect to $N=\ln(a)$.
We can obtain the evolution of the density parameters by means of the following relations:
\begin{eqnarray}
    \Omega_{\Phi}&=&xy+uv, \ b^2=\Omega_b,\ \Delta^2=\Omega_{\Lambda}, \ r^2=\Omega_{r}.
\end{eqnarray}
In \cite{Matos2009}, it was shown that $s'=s_1s^{-k}$, where $s_1$ is a constant to determine. Then we can solve the system of equations \eqref{ec:system} numerically using that result for $s'$, and the constraint $xy+uv+b^2+\triangle^2+r^2=1$. Once we solve the system \eqref{ec:system} numerically, we can obtain ${\theta}_0',{R}_0'$ and $R_0$ from the following relations:
\begin{eqnarray}
    \theta_0'^2&=&-\frac{s^2}{4}\left(\frac{x}{u}-\frac{y}{v}\right)^2,\nonumber\\
    R_0'&=&\frac{\sqrt{3uv}}{2\kappa}\left(\frac{x}{u}+\frac{y}{v}\right),\nonumber\\
    R_0&=&\frac{\sqrt{3uv}}{\kappa s}.
\end{eqnarray}
Finally, we solve the following differential equation system for $T_R$:
\begin{eqnarray}
\label{eq:TRsystem}
    T_R''+\left(3-\frac{3}{2}\frac{\Pi}{a}+2\theta_0'^2\frac{R_0}{R_0'}\right)T_R'+\left(\frac{l^2s^2}{a^2m^2}+s^2-3\theta_0'^2\right)T_R=0,\nonumber\\
    T_R'-\frac{qT_0s}{am}\frac{R_0'}{\theta_0'}+\frac{R_0'}{R_0}T_{R}=0,
\end{eqnarray}
where we have assumed $\lambda\approx0$.

\begin{figure}[h]
    \centering
     \includegraphics[width=10cm]{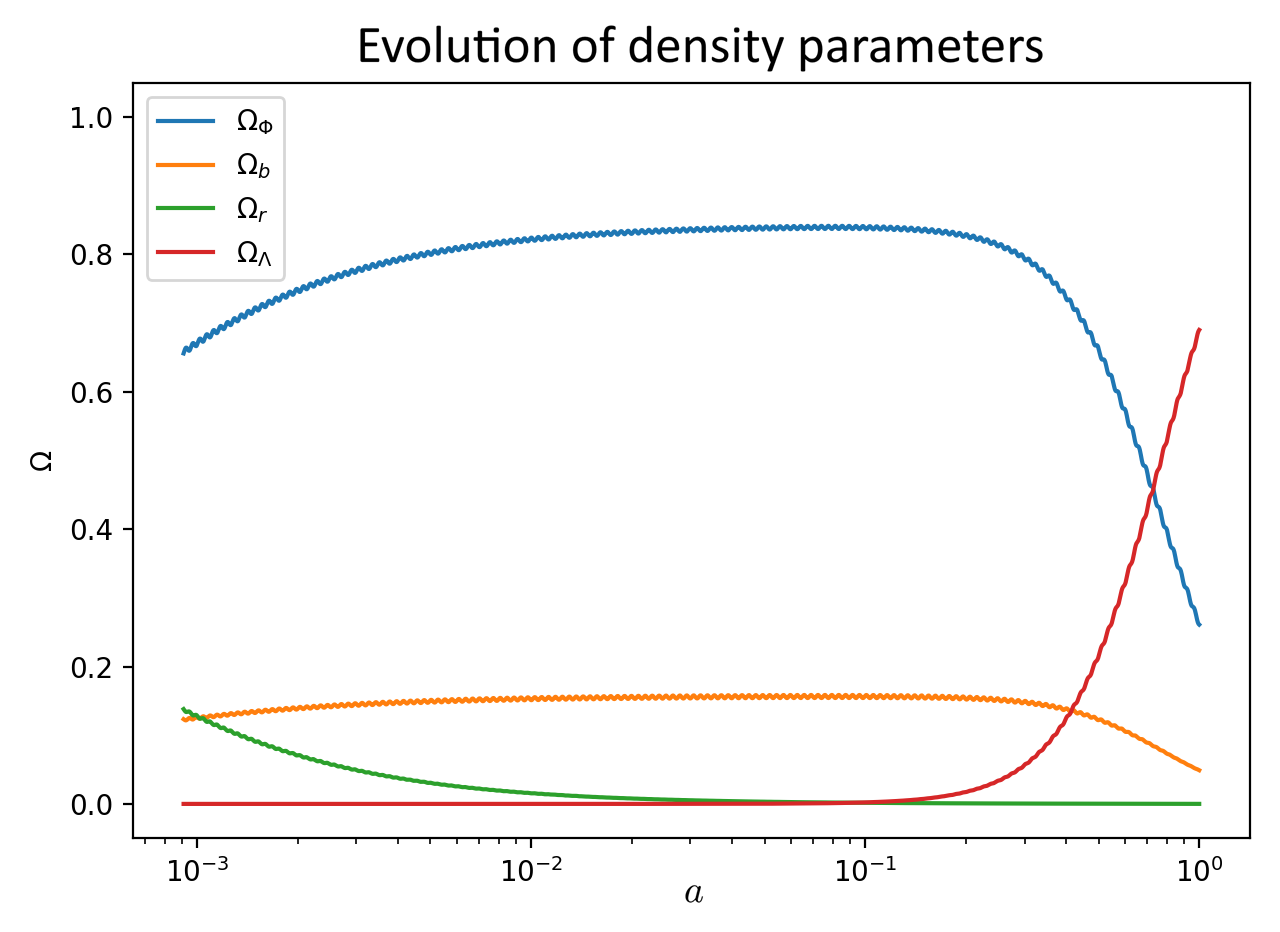}
    \caption{Evolution of the density parameters with respect to the scale factor, $a$. $\Omega_{\Phi}$ is the density parameter of the SFDM.}
\label{density-parameters}
\end{figure}

The background evolution obtained from this system reproduces the standard sequence of radiation, matter, and dark-energy domination, while allowing the SFDM component to form halos with the multistate structure discussed in \cite{bernal2025natural} (see Figure~\ref{density-parameters}). This guarantees that the mechanism proposed here for the generation of galactic magnetic fields is fully consistent with the cosmological evolution of the scalar field.

\subsection{Spatial distribution of SFDM}
\label{subsec:spatialdis}

The density of the fluctuation that collapses to form the structure is: 
\begin{equation}
    n=m_{\Phi}\delta\Phi\delta\Phi^* \approx m_{\Phi}(\delta R^2+R_0^2\delta\theta^2) \approx m_{\Phi}\delta R^2,
\end{equation}
where $n$ is the number density. Then the mass density is given by: $\rho_{\Phi}=m_{\Phi}^2\delta\Phi\delta\Phi^*=m_{\Phi}^2\delta R^2$,
and the  total mass in a SFDM halo is given by: 
\begin{eqnarray}
    M&=&m_{\Phi}^2\int\delta R^2r^2\sin\theta drd\theta d\varphi=m_{\Phi}^2\int\sum_{k,k'}T_R^2A_{0k}j_k(lr)A_{0k'}j_{k'}(lr)Y_{k}^0(\theta,\varphi)Y_{k'}^0(\theta,\varphi)r^2\sin\theta dr d\theta d\varphi\nonumber\\
    &=&m_{\Phi}^2T_R^2\sum_k\int A_{0k}^2j_k(lr)^2r^2dr.
\end{eqnarray}
We now define $\rho_k(r,\eta)=\rho_{0k}j_k(lr)^2$ and $\rho_{0k}=m_{\Phi}^2A_{0k}^2T_R^2(\eta)$, where $\rho_{0k}$ is the central density for the galaxy for the $k$-state. 
Then, for the ground and first excited states, we have: 
\begin{eqnarray}
    M_0(r,\eta)&=&\frac{2\pi\rho_0(r,\eta)}{l^3}(lr-\cos(lr)\sin(lr)),\nonumber\\
    M_1(r,\eta)&=&\frac{2\pi\rho_1(r,\eta)}{l^3}\left(\frac{\cos(2lr)}{lr}-\frac{1}{lr}+\cos(lr)\sin(lr)+lr\right).
\end{eqnarray}
Therefore, the spatial distribution of SFDM is the same as in \cite{bernal2025natural}, where there is no gauge field. The main difference is that, in this case, the mass varies with time, which depends on the evolution of $T_R(\eta)$. The current value is obtained by evaluating at $N=0$ or $a=1.$ After using the densities $\rho_0$ and $\rho_1$ found in \cite{bernal2025natural} for the Milky Way, Andromeda and Centaurus A, we can find the values for $m_{\Phi}A_{00}$, $m_\Phi A_{01}$, and $T_R(N=0)\sim 10^{6}$ for these three galaxies. With these values, we obtain the rotation curves for such galaxies, confirming that the rotation curves are not affected by the presence of the gauge field. The values we use are presented in Table \ref{tab:central}, and the rotation curves are the same as reported in \cite{bernal2025natural}.

This result shows that the gauge field $B_\mu$, and hence the magnetic field it generates, acts as a ``spectator'' with respect to the spatial distribution of SFDM. The halo preserves the same core-excited-state structure used to explain the VPOS in \cite{bernal2025natural}, while simultaneously sourcing a non-trivial large-scale magnetic field.

\begin{table}[h]
    \centering
         \begin{tabular}{|c|c|c|c|c|c|}
        \hline    
        Galaxy & $\rho_0 $& $\rho_1$ & $l$ & $m_{\Phi}A_{00}$ & $m_{\Phi}A_{01}$\\
        & $(10^{-2}M_{\odot}\text{pc}^{-3})$ & $(10^{-2}M_{\odot}\text{pc}^{-3})$ & (kpc$^{-1}$) & (eV$^2$)&(eV$^2$)\\
        \hline
        Milky Way & 
        3.88 & 3.19 & 0.23 & $3.3\times 10^{-9}$ &
        $1.4\times10^{-9}$ \\
        Andromeda &
        3.46 & 0.766 & 0.169 & $3.3\times 10^{-9}$ & $1.4\times 10^{-9}$ \\
        Centaurus A & 
        750.77 & 356.31 & 2.09 & $4.76\times 10^{-8}$&$3.22\times 10 ^{-8}$ \\
        \hline
    \end{tabular}
       \caption{Values found for $m_{\Phi}A_{00}$ and $m_{\Phi}A_{01}$. The values for $\rho_0$, $\rho_1$, for the ground and excited states of the scalar field, respectively, and $l$, are taken from the previous work by \cite{bernal2025natural}, after fitting the rotation curves of such galaxies.}
 \label{tab:central}
\end{table}

\subsection{Gravitational Potential}
\label{sec:potential}

In this subsection, we compute the Newtonian potential $\phi$ from the component $\delta G_{0}^0=\kappa^2\delta T_0^0$ of the field equations. We obtain: 
\begin{equation}
    2\nabla^2\phi-6\mathcal{H}(\dot{\phi}+\mathcal{H}\phi)=\kappa^2\left[\delta\dot{\Phi}\dot{\Phi}_0^{*}+\dot{\Phi}_0\delta\dot{\Phi}^{*}+a^2\delta V-2\phi\dot{\Phi}\dot{\Phi}^*\right]
    +iq\kappa^2\left(\Phi_0\dot{\Phi}_0^{*}-\Phi_0^{*}\dot{\Phi}_0\right)B_0, 
\end{equation}
where $\delta V=V_{,\Phi_0}\delta\Phi+V_{,\Phi_0^{*}}\delta\Phi^{*}$. Then, in terms of the Mandelung transformation, this equation can be rewritten as: 
\begin{equation}
 \nabla^2\phi-3\mathcal{H}\left(\dot{\phi}+\mathcal{H}\phi\right)=\kappa^2\left[\dot{\theta}_0^2R_0\delta R+\dot{R}_0\delta\dot{R}+m^2a^2R_0\delta R-\phi(\dot{R}_0^2+R_0^2\dot{\theta}_0^2)+qB_0R_0^2\dot{\theta}_0\right].
\end{equation}
We can see from the above equation that the potential field is affected only by the time component of the gauge field. After using equation \eqref{eq:B0}, we find the following differential equation for $\phi$:
\begin{equation}
    \nabla^2\phi -\kappa^2\delta\dot{R}\left(\dot{R}_0-\frac{R_0\dot{\theta}_0^2}{\dot{R}_0}\right)-\kappa^2\delta R(2\dot{\theta}_0^2R_0+m^2a^2R_0)+\phi(\dot{R}_0^2+\dot{\theta}_0^2R_0^2)=0.
\end{equation}
We can solve this equation by assuming that $\phi=T_p(\eta)Y_k^m(\theta,\varphi)p(r)$, and using the fact that $\delta R=j_k(lr)Y_k(\theta,\varphi)T_R(\eta)$. We then have:
\begin{equation}
    \frac{1}{r^2}\frac{\partial}{\partial r}\left(r^2\frac{\partial p}{\partial r}\right)+\left(\Omega_1^2-\frac{k(k+1)}{r^2}\right)p=\kappa^2j_k(lr)\left[\frac{\dot{T}_R}{T_p}\left({\dot{R}_0}-\frac{R_0^2\dot{\theta}_0^2}{\dot{R}_0}\right)+\frac{T_R}{T_p}\left(2\dot{\theta}_0^2R_0+m^2a^2R_0\right)\right],
\end{equation}
where $\Omega_1^2=\kappa^2\left(\dot{R}_0^2+R_0^2\dot{\theta}_0^2\right)$, whose solution is given by:
\begin{eqnarray}
    p(r) = \sum_k j_k(\Omega_1 r)+\frac{\Omega_2j_k(lr)}{\Omega_1^2-l^2},
\end{eqnarray}
with $\Omega_2=\left[\frac{\dot{T}_R}{T_p}\left({\dot{R}_0}-\frac{R_0^2\dot{\theta}_0^2}{\dot{R}_0}\right)+\frac{T_R}{T_p}\left(2\dot{\theta}_0^2R_0+m^2a^2R_0\right)\right]$. The potential $p$ has the same functional form as the potential found in \cite{bernal2025natural}, without the gauge field.

\section{Results for the magnetic field}
\label{sec:results}

From equations \eqref{eq:Bx}-\eqref{eq:Bz}, we can see that the strength of the magnetic fields depends on the value of the constants $c_x$, $c_y$, $c_z$, $A_{00}$, $T_0$, and $l$. Also, we can see that these fields vary with respect to the cosmic time, and this variation is proportional to $\mathcal{H}/a^3=H/a^2$, and its current value is obtained at $a=1$.

\begin{figure}
    \centering
    \includegraphics[width=8cm]{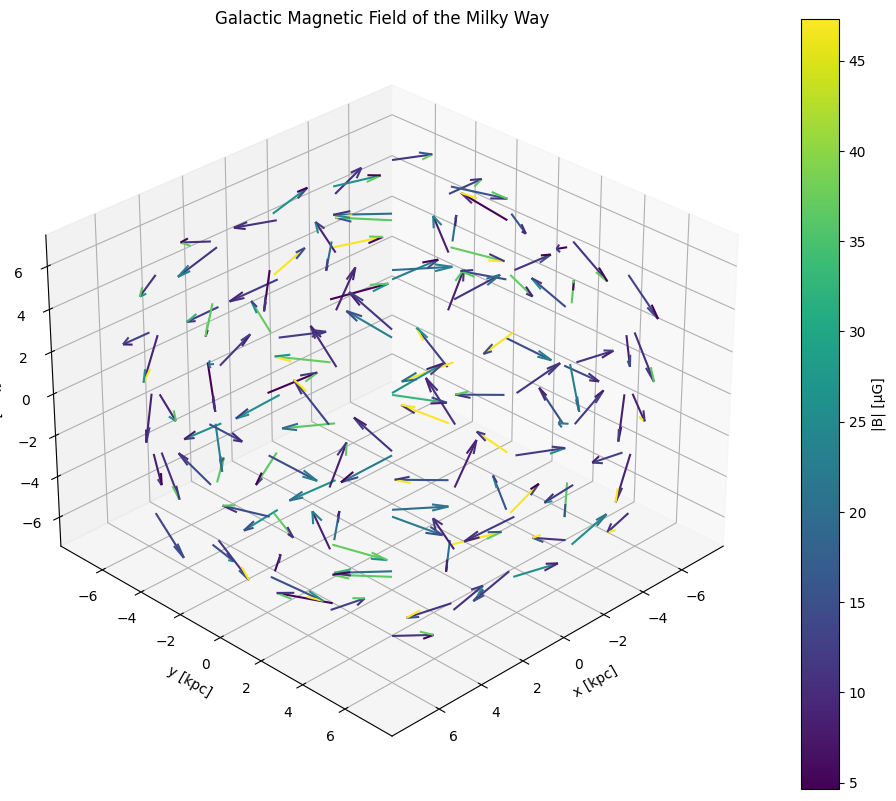}\includegraphics[width=8cm]{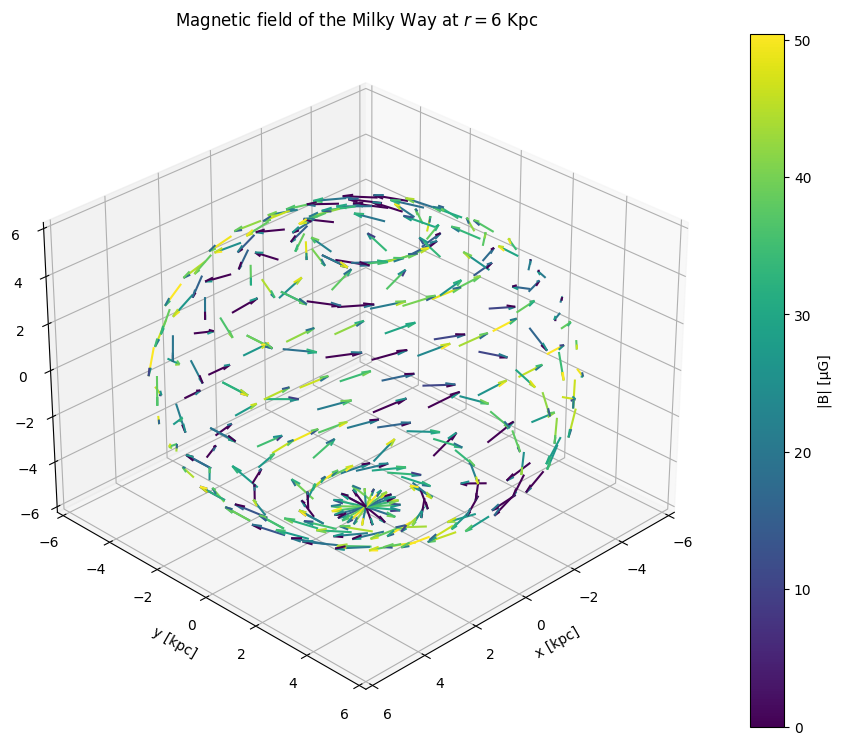}\\
    \includegraphics[width=8cm]{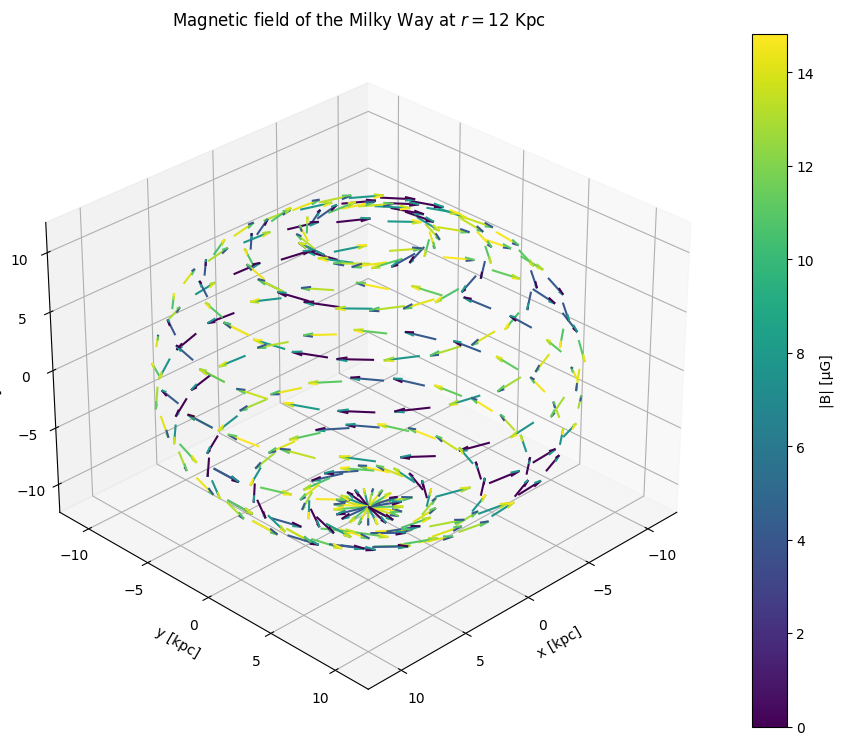}\includegraphics[width=8cm]{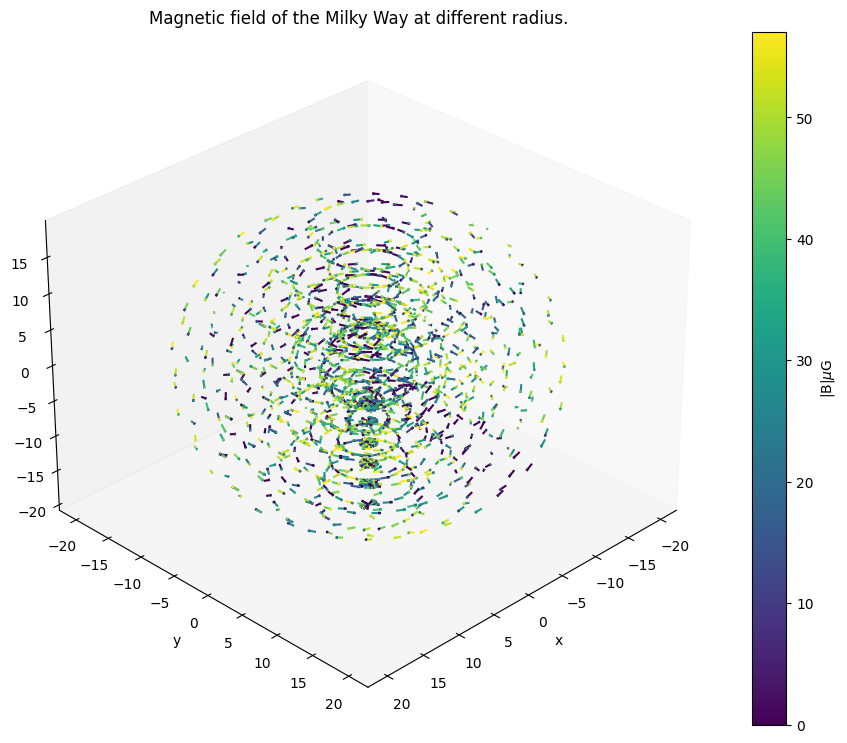}
    \caption{Galactic magnetic field of the Milky Way generated by the charged multistate SFDM halo at the present epoch ($a=1$), shown on physical galactic scales. Upper left panel: The magnetic field is sampled on a Cartesian grid within a cubic domain of side $16$ kpc (a volume of 4096 kpc$^3$) and only points satisfying $5\leq r\leq10$ kpc are displayed. Upper right panel: magnetic field configuration at a fixed radius of $6$~kpc. Bottom left panel: magnetic field at a radius of $12$~kpc. Bottom right panel: magnetic field structure at different galactocentric radii. The field amplitude is expressed in microgauss units, allowing for direct comparison with observational estimates. The spatial coherence over kiloparsec scales reflects the underlying spherical Bessel modes of the scalar-field perturbations, while the angular structure follows from the spherical harmonic decomposition. In particular, the last panel exhibits a dipolar-like pattern, consistent with the two-lobed structure associated with the first excited state of the scalar field. For parameter values compatible with realistic halo sizes and rotation curves, the magnetic field naturally reaches amplitudes of the order of microgauss, without requiring additional amplification mechanisms.
    }
    \label{fig:MilkyWay}
\end{figure}

\begin{figure}
    \centering   
    \includegraphics[width=8cm]{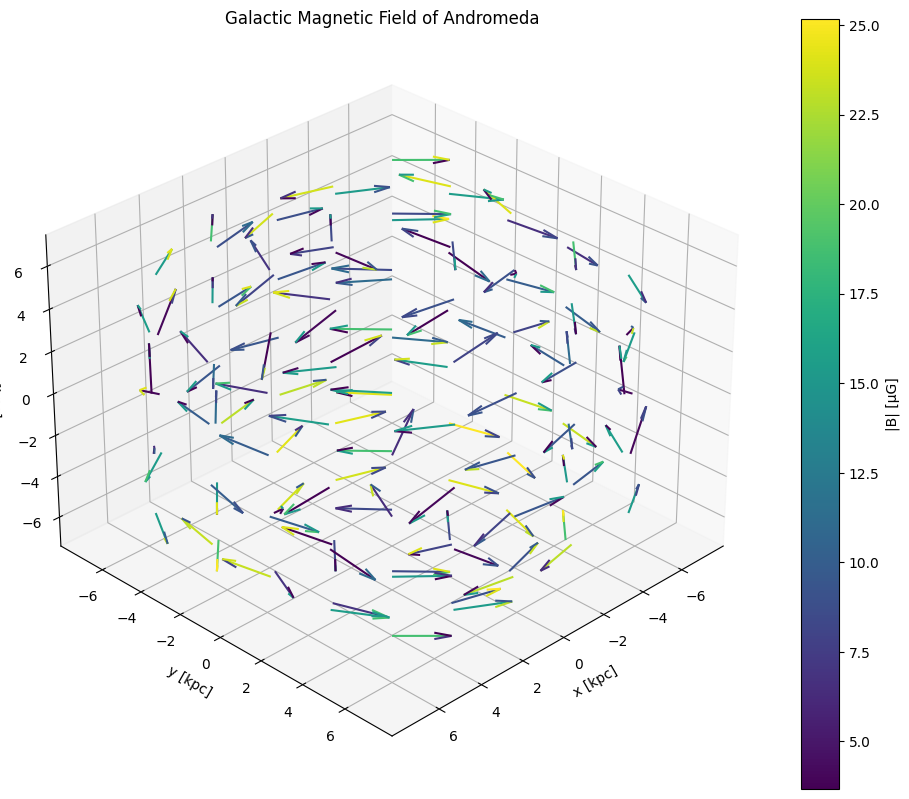}\includegraphics[width=8cm]{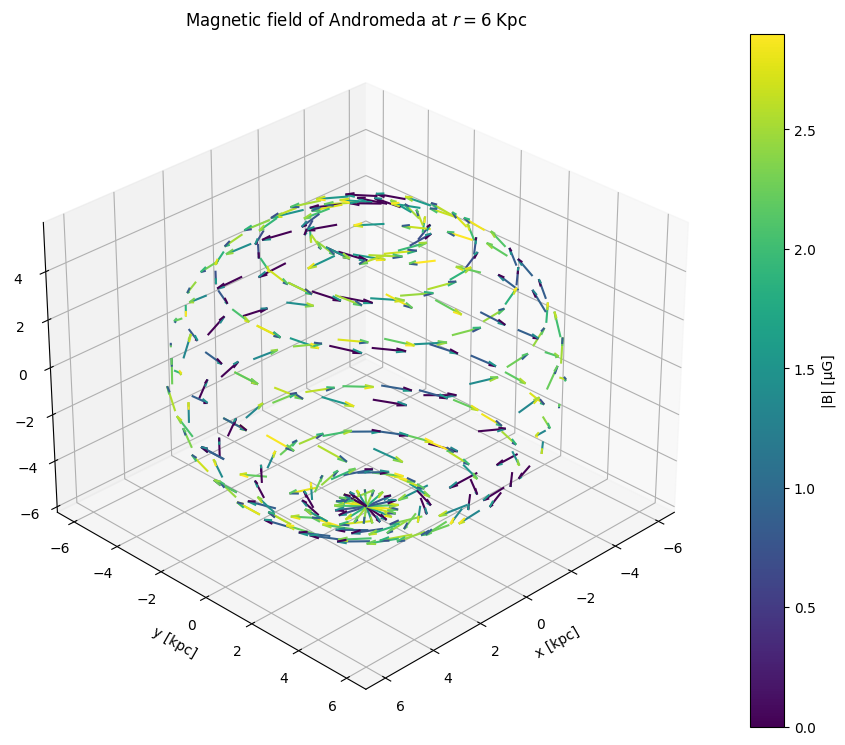}\\
    \includegraphics[width=8cm]{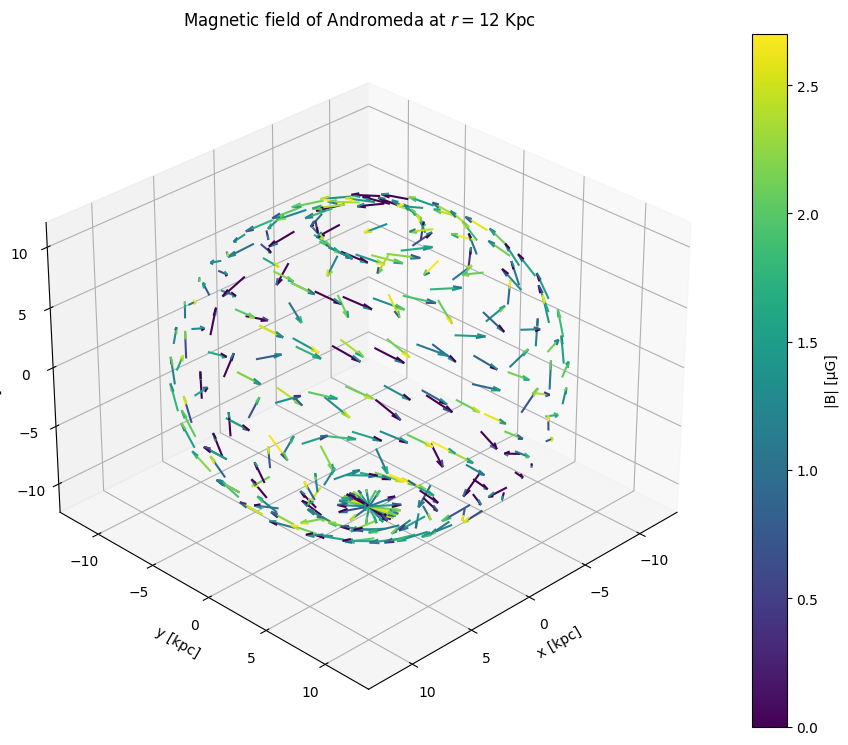}\includegraphics[width=8 cm]{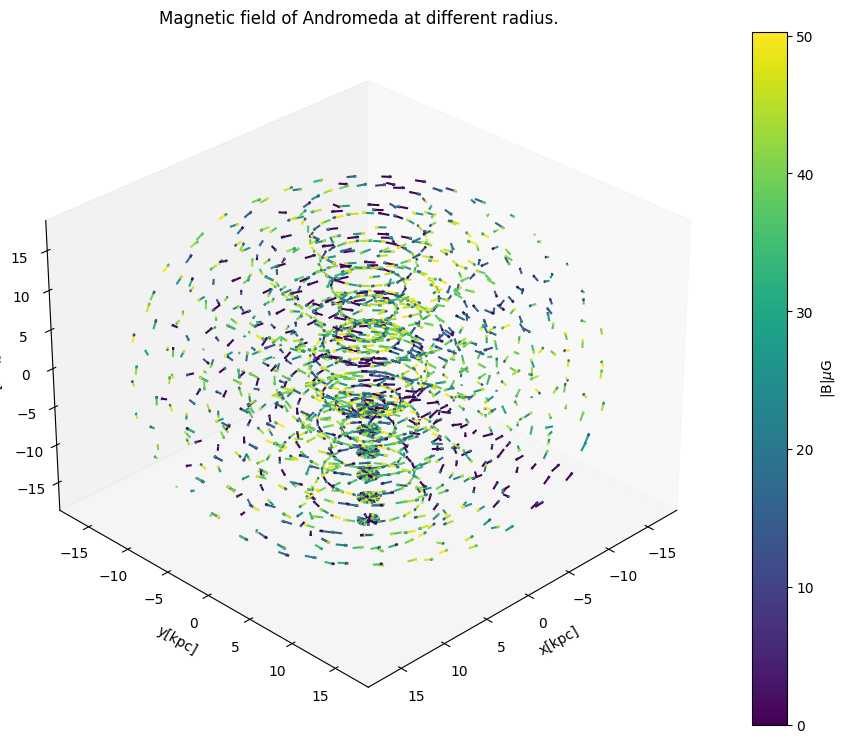}
    \caption{Galactic magnetic field of Andromeda generated by the charged multistate SFDM halo at the present epoch ($a=1$). Upper left side: The magnetic field is sampled on a Cartesian grid within a cubic domain of side $16$ kpc (a volume of 4096 kpc$^3$) and only points satisfying $5\leq r\leq10$ kpc are displayed. Upper right side: Galactic magnetic field at a radius of $6$ Kpc. Bottom left side: Magnetic field of Andromeda at a radius of $12$ kpc. Bottom right side: Magnetic field of Andromeda at different radii. This last panel exhibits a dipolar-like pattern, consistent with the two-lobed structure associated with the first excited state of the scalar field.}
    \label{fig:Andromeda}
\end{figure}

\begin{figure}
    \centering    \includegraphics[width=8cm]{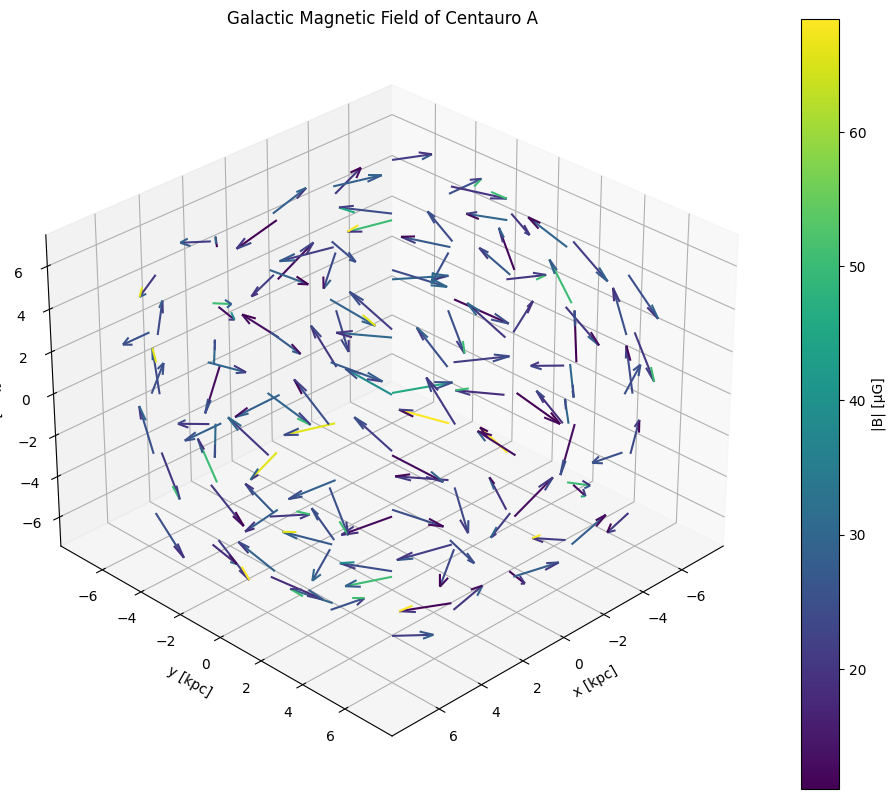}\includegraphics[width=8cm]{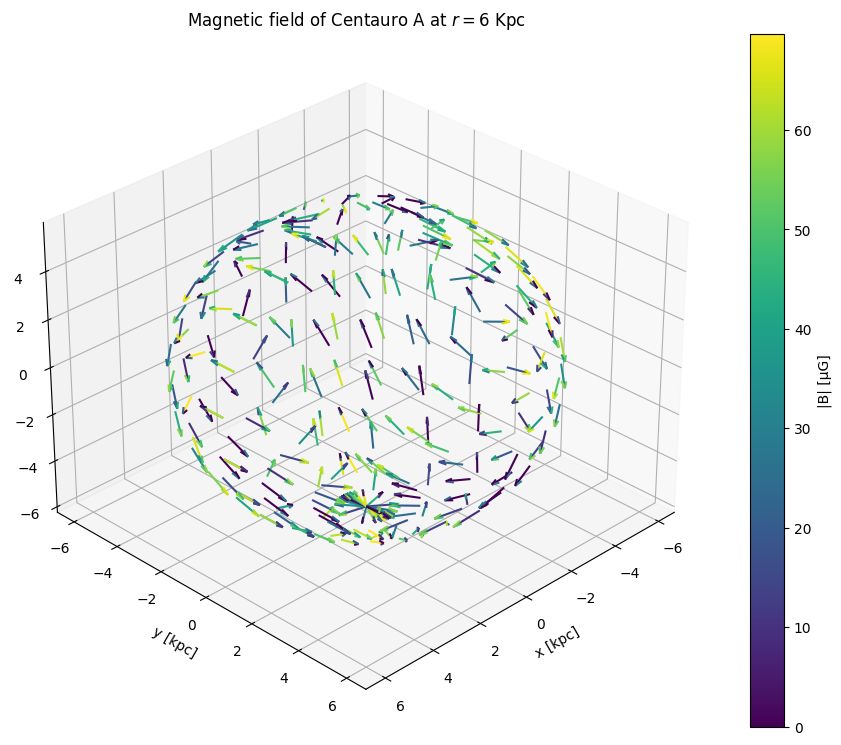}\\
    \includegraphics[width=8cm]{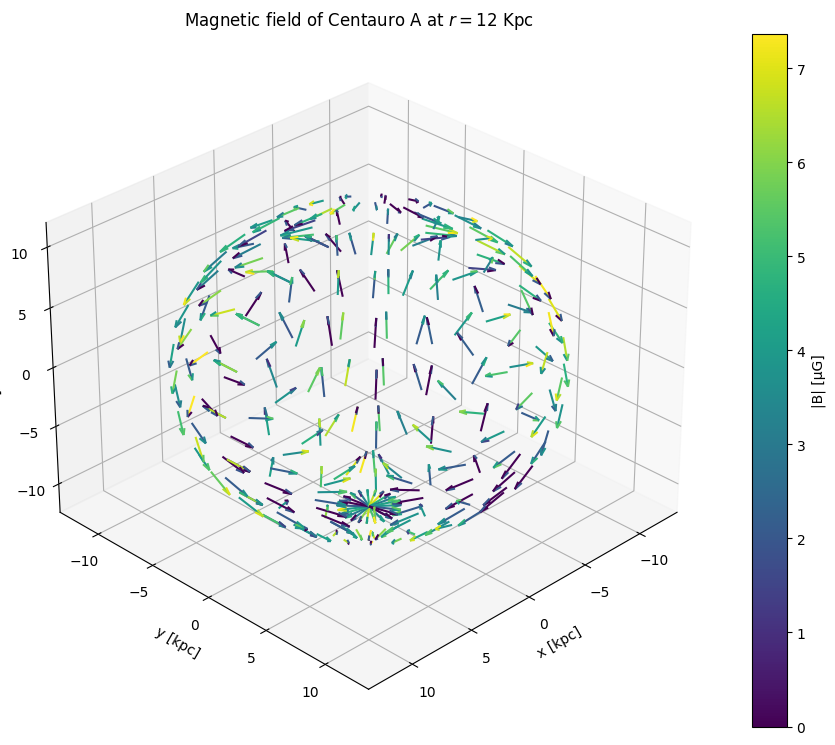}\includegraphics[width=8 cm]{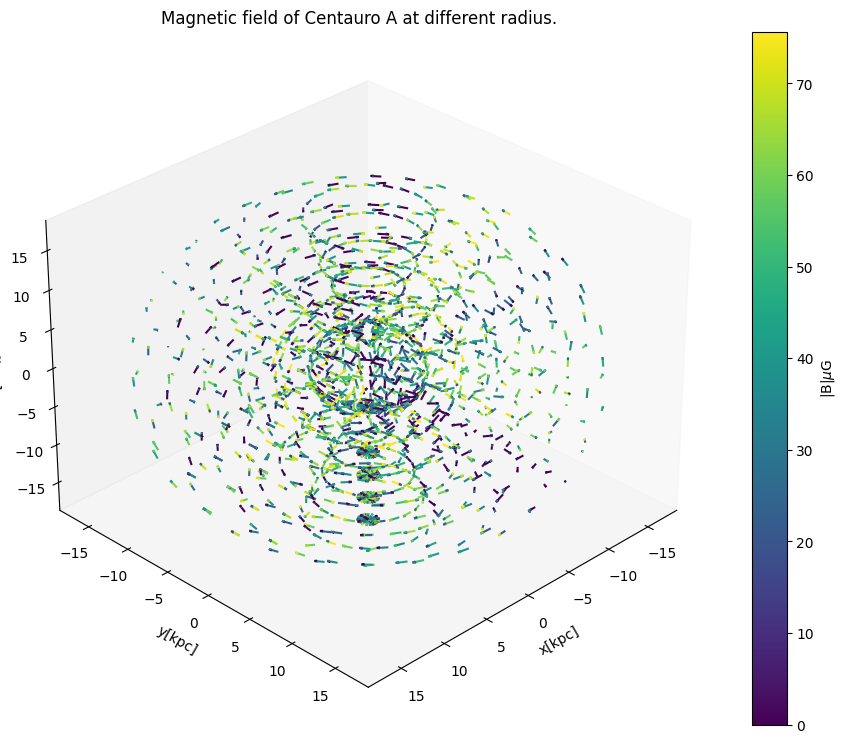}
    \caption{Galactic magnetic field of Centaurus A generated by the charged multistate SFDM halo at the present epoch ($a=1$). Upper left side: The magnetic field is sampled on a Cartesian grid within a cubic domain of side $16$ kpc (a volume of 4096 kpc$^3$) and only points satisfying $5\leq r\leq10$ kpc are displayed. Upper right side: Galactic magnetic field at a radius of $6$ kpc. Bottom left side: Magnetic field of Centaurus A at a radius of $12$ kpc. Bottom right side: Magnetic field of Centaurus A at different radii. This last panel exhibits a dipolar-like pattern, consistent with the two-lobed structure associated with the first excited state of the scalar field.}
    \label{fig:CentauroA}
\end{figure}

In Figure \ref{fig:MilkyWay}, we plot the magnetic fields for the Milky Way; in Figure \ref{fig:Andromeda}, the resulting magnetic fields for 
Andromeda, and in Figure \ref{fig:CentauroA} the results for Centaurus A, all the fields at the present time, $a=1$. The values that we use for the constants are shown in Table \ref{tab:valuesmagnetic}; these values were found using the differential evolution algorithm for global optimization \cite{Storn:1997uea,2020SciPy}, included in the \texttt{scipy.optimize} library of Python. For $l$, we use the values found in \cite{bernal2025natural}, shown in Table \ref{tab:central}.

\begin{table}[h]
    \centering
    \resizebox{\textwidth}{!}{%
    \begin{tabular}{|c|c|c|c|c|c|c|c|}
        \hline    
        Galaxy & $\Omega_x$& $\Omega_y$ & $\Omega_z$ & $\mathcal{H}_0A_{00}T_0c_x$ & $\mathcal{H}_0A_{00}T_0c_y$ & $\mathcal{H}_0A_{00}T_0c_z$ & $\mathcal{H}_0T_0A_{00}$\\
        & ($10^{11}$eV) & ($10^{11}$eV) & ($10^{11}$eV) & ($10^{20}$eV) & ($10^{20}$eV) & ($10^{20}$eV) &($10^{-30}$eV$^2$)\\
        \hline
        Milky Way & 
        1.83 & 1.72 & 0.671 & 2.146 & 3.001 & 9.725  &  3.3\\  
        Andromeda& 1.15  & 3.28  & 6.40 & 0.809 & 3.271 & 15.97 & 3.3 \\
        Centaurus A & 5.64 & 3.55 & 6.09 & 21.20 & 8.284 & 20.21 & 46 \\
        \hline
    \end{tabular}}
    \caption{Values of the parameters found using the differential evolution algorithm for global optimization, included in the \texttt{scipy.optimize} library of Python. Here $\mathcal{H}_0$, is the current value of $\mathcal{H}$ (at $a=1$).}
\label{tab:valuesmagnetic}
\end{table}

We can see from Figure \ref{fig:MilkyWay} that the galactic magnetic fields for the Milky Way are rotating around the $z$ axis, and that the direction of the rotation depends on the radial distance to the center of the Galaxy; the magnetic field strength increases toward the Galactic center. We have the same behavior for Andromeda (Figure \ref{fig:Andromeda}), and for Centarus A the rotation is perpendicular to the $z$ axis (Figure \ref{fig:CentauroA}).

The three Figures show also that the magnetic field generated by the multistate SFDM halo naturally reaches amplitudes of the order of microgauss at the present epoch ($a=1$), spanning radial distances of a few kiloparsecs, which correspond to the region where large-scale coherent magnetic fields are observationally inferred in spiral galaxies, without the need for additional amplification mechanisms. For parameter values consistent with realistic halo sizes and rotation curves \cite[reported in][]{bernal2025natural}, we obtain magnetic fields of the order of microgauss. This result suggests that galactic magnetic fields can emerge as a direct consequence of the charged multistate SFDM configuration, rather than requiring a primordial seed field or a highly efficient dynamo process.

Furthermore, in the last panel of the three figures, a dipolar-like structure emerges as a result of the two-lobed morphology associated with the first excited state of the scalar field, highlighting the connection between the multistate SFDM halo that can explain the VPOS structure of the Milky Way, Andromeda and Centaurus A, and the large-scale organization of the magnetic field.

\section{Conclusions}
\label{sec:conclusions}
In order to explain the large-scale magnetic fields of the order of microgauss present in galaxies, we follow the idea introduced in \cite{hernandez2019could}. We assume that dark matter is a complex scalar field endowed with an internal $U(1)$ symmetry. The corresponding Lagrangian includes a new charge $q$, associated with the gauge field $B_{\mu}$. Two possibilities must be studied. The first is that the gauge field corresponds to the standard model photon in which case the charge must be extremely small. The second possibility is that the gauge boson is a dark photon that interacts only gravitationally with standard model particles and through a mixing kinetic term with the photon of the standard model.

In this work, we focus on the first scenario and assume that the charge $q$ is very small. Then we find that the presence of this gauge field does not modify the distribution of the multistate SFDM found in \cite{bernal2025natural}. Furthermore, we find analytical expressions in terms of Bessel functions of the temporal and spatial components of this gauge field, and with these results, we compute the electric and magnetic fields generated by this gauge field. We found that we can obtain magnetic fields with strengths of the order of microgauss, and that their magnitude depends only on the values of the free parameters of the model. We find the value of these free parameters for 3 different galaxies: Milky Way, Andromeda, and Centaurus A, and plot their magnetic fields at the present time, $a=1$. We found that the direction of the magnetic fields depends on the radial distance to the galactic center, and the magnetic field strength increases toward the Galactic center.

Thus, we have shown that a charged multistate SFDM halo can naturally generate large-scale magnetic fields whose spatial structure is governed by the same spherical Bessel and harmonic modes that describe the ground and first excited states of the halo. The gauge field inherits the ``atomic'' structure of the SFDM configuration, leading to coherent magnetic fields that trace the dipolar-like excited modes of the scalar field without altering the underlying dark-matter density profile.

This result establishes a close connection between two apparently different galactic-scale phenomena. On the one hand, the multistate SFDM model explains the anisotropic distribution of satellite galaxies (VPOS) as a consequence of the excited states of the halo \cite{bernal2025natural} in the 3 galaxies: Milky Way, Andromeda, and Centaurus A. On the other hand, when the scalar field carries an extremely small electric charge, the same excited states act as sources for a large-scale magnetic field. In this unified picture, both the VPOS and galactic magnetic fields emerge as manifestations of the quantum nature of the dark-matter halo.


\section*{Acknowledgements}
MH-M acknowledges financial
support from SECIHTI postdoctoral fellowships.
This work was partially supported by SECIHTI M\'exico under grants CBF-2025-G-1720 and CBF-2025-G-176. Also  by the grant I0101/131/07 C-234/07 of the Instituto Avanzado de Cosmolog\'ia (IAC) collaboration (http://www.iac.edu.mx/), and for the computing time granted by LANCAD and SECIHTI in the Supercomputer Hybrid Cluster "Xiuhcoatl" at GENERAL COORDINATION OF INFORMATION AND COMMUNICATIONS TECHNOLOGIES (CGSTIC) of CINVESTAV, URL: http://clusterhibrido.cinvestav.mx.  M.A. also acknowledges financial support from DGAPA-UNAM project IN100523.

\bibliographystyle{JHEP}
\bibliography{bibliography.bib}

\end{document}